\documentclass[aps,prx,10pt,superscriptaddress,twocolumn,longbibliography,notitlepage,nofootinbib]{revtex4-2}
\usepackage[utf8]{inputenc}
\usepackage[T1]{fontenc}

\usepackage{amsmath}
\usepackage{amsthm}
\usepackage{amsfonts}
\usepackage{amssymb}
\usepackage{bm}
\usepackage{bbm}
\usepackage{bbold}
\usepackage{physics}
\usepackage{mathtools}

\usepackage{xcolor}
\definecolor{color1}{rgb}{0,0,0.7}
\definecolor{color2}{rgb}{0.85,0,0}

\usepackage[breaklinks=true,colorlinks,citecolor=red,linkcolor=color2,urlcolor=color1]{hyperref}

\newcommand{\eref}[1]{\textcolor{color2}{\hyperref[#1]{eq.~(\ref{#1})}}}
\newcommand{\Eref}[1]{\textcolor{color2}{\hyperref[#1]{Eq.~(\ref{#1})}}}
\newcommand{\fref}[1]{\textcolor{color2}{\hyperref[#1]{Fig.~\bfseries{\ref{#1}}}}}
\newcommand{\sref}[1]{\textcolor{color2}{\hyperref[#1]{Sec.~\bfseries\ref{#1}}}}
\newcommand{\eq}[1]{\textcolor{color2}{(\ref{#1})}}
\newcommand{\aref}[1]{\textcolor{color2}{\hyperref[#1]{App.~\bfseries\ref{#1}}}}
\newcommand{\stepref}[1]{\textcolor{color2}{\hyperref[#1]{Step~\bfseries\ref{#1}}}}

\newcommand{\W}{\mathcal{W}}
\newcommand{\Q}{\mathcal{Q}}
\newcommand{\B}{\mathcal{B}}
\newcommand{\Wir}{W_{\rm diss}}

\usepackage{titlesec}
\titlespacing*{\section}{0pt}{10pt plus 4pt minus 5pt}{7pt plus 3pt minus 5pt}
\titlespacing*{\subsection}{0pt}{5pt plus 3pt minus 1pt}{4pt plus 2pt minus 4pt}
\titleformat{\section}{\centering\bfseries}{\thesection.}{.5em}{}

\begin{document}

\title{Energy-Time-Accuracy Tradeoffs in Thermodynamic Computing}

\author{Alberto Rolandi}
\affiliation{Atominstitut, TU Wien, 1020 Vienna, Austria}
\affiliation{Institute for Quantum Optics and Quantum Information - IQOQI Vienna, Austrian Academy of Sciences, Boltzmanngasse 3, A-1090 Vienna, Austria}

\author{Paolo Abiuso}
\affiliation{Institute for Quantum Optics and Quantum Information - IQOQI Vienna,
Austrian Academy of Sciences, Boltzmanngasse 3, A-1090 Vienna, Austria}

\author{Patryk Lipka-Bartosik}
\affiliation{Center for Theoretical Physics, Polish Academy of Sciences, Warsaw, Poland}
\address{Institute of Theoretical Physics, Jagiellonian University, 30-348 Kraków, Poland}

\author{Maxwell Aifer}
\affiliation{Normal Computing Corporation, New York, New York, USA}


\author{ Patrick J. Coles}
\email{patrick@normalcomputing.ai}
\affiliation{Normal Computing Corporation, New York, New York, USA}

\author{Martí Perarnau-Llobet}
\email{marti.perarnau@uab.cat}
\affiliation{F\'isica Te\`orica: Informaci\'o i Fen\`omens Qu\`antics, Department de F\'isica, Universitat Aut\`onoma de Barcelona, 08193 Bellaterra (Barcelona), Spain}

\begin{abstract}
In the paradigm of thermodynamic computing, instead of behaving deterministically, hardware undergoes a stochastic process in order to sample from a distribution of interest. While it has been hypothesized that thermodynamic computers may achieve better energy efficiency and performance, a theoretical characterization of the resource cost of thermodynamic computations is still lacking.
Here, we analyze the fundamental trade-offs between computational accuracy, energy dissipation, and time in  thermodynamic computing. Using geometric bounds on entropy production, we derive general limits on the energy–delay–deficiency product (EDDP), a stochastic generalization of the traditional energy–delay product (EDP). While these limits can in principle be saturated, the corresponding optimal driving protocols require full knowledge of the final equilibrium distribution, i.e., the solution itself. To overcome this limitation, we develop quasi-optimal control schemes that require no prior information of the solution and demonstrate their performance for matrix inversion in overdamped quadratic systems.  
The derived bounds extend beyond this setting to more general potentials, being directly relevant to recent proposals based on non-equilibrium Langevin dynamics. 
\end{abstract}

\maketitle

\begin{figure*}[ht]
    \centering
    \includegraphics[width=.85\textwidth]{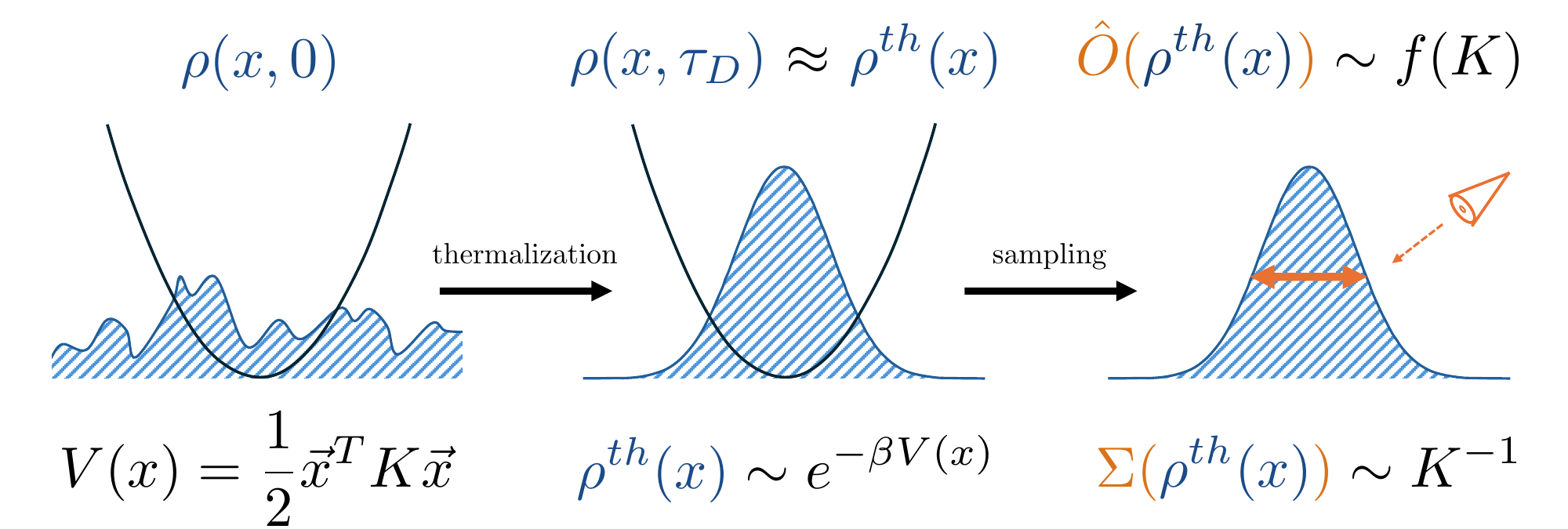}
    \vspace*{-5pt}
    \caption{{\bfseries Thermodynamic algorithm for computing $f(K)$.} In this illustration, the probability density of the state (in blue) is in some arbitrary initial distribution when the potential $V(x) = \frac{1}{2}\vec{x}^TK\vec{x}$ (in black) is applied. The probability density of the state naturally evolves towards thermal equilibrium (\stepref{s1}), after which the observer can start sampling properties of the system to directly measure $f(K)$ (\stepref{s2}). For example, the correlation matrix $\Sigma$ of the positions in the thermal distribution corresponds to the inverse of $K$.}
    \vspace*{-15pt}
    \label{fig:thermo_alg}
\end{figure*}
\section{Introduction}

Computational time, accuracy, and system size have long guided progress in computational architectures, algorithms, and hardware. With the rapid expansion of AI technologies, however, power consumption is becoming a central bottleneck in modern computing devices. A basic metric to quantify the balance between power  and computational speed is the energy–delay product (EDP), defined as the product of the energy consumed and the execution time of a computation~\cite{burr1991ultra}. Clarifying the fundamental limits of EDP is crucial not only for optimizing current platforms, but also for assessing new physics-based approaches~\cite{Markovi2020}, such as thermodynamic computing~\cite{Conte2019Thermodynamic,coles2023thermodynamic_published}, which aims to exploit physical principles to achieve ultra-low-energy computation.



The limits of EDP go back to Landauer's principle, which establishes a minimum energy cost for irreversible computations. Specifically, erasing one bit of information dissipates at least $k_{\rm B} T \ln(2)$ Joules, where $k_{\rm B}$ is the Boltzmann constant and $T$ is the system temperature~\cite{landauer1961irreversibility}. 
Real implementations have  substantial overheads, such as the energetic cost of precise timing control~\cite{friedman2002clock} and voltage switching~\cite{chandrakasan1992low}, leakage currents arising due to scaling~\cite{roy2003leakage}, or the cost of error data-memory movements~\cite{horowitz20141}. 
Together, these contributions typically raise energy costs of computation by many orders of magnitude above the Landauer value. Even in the absence of such practical limitations, achieving Landauer’s limit requires the system to remain in equilibrium at all times, so any (ideal) computation performed in finite time will necessarily lie above Landauer’s limit.
To address this gap, recent advances in stochastic thermodynamics~\cite{Wolpert2024,Chattopadhyay2025,chattopadhyay2025understanding} have introduced finite-time extensions of Landauer’s bound~\cite{proesmans2020finite,Van_Vu_2022,Rolandi_2023,Taranto2023Landauer} and characterized the energetics of minimal models of logic gates~\cite{wolpert2020thermodynamics,freitas2021stochastic,Gao2021,Konopik2023,meier2024autonomous,AuffevesPRX2022,PhysRevX.14.021026,Van_Vu_2025}, both revealing that faster computations (i.e. shorter delays) entail higher energy costs. Furthermore, proof-of-principle experimental demonstrations  of Landauer's principle  in colloidal particles, nanomagnetic systems and quantum molecular magnets, have been achieved~\cite{berut2012experimental,jun2014high,Hong2016,Martini2016,Gaudenzi2018,scandi2022constant}.

Recent advances have further characterized EDP across diverse platforms.  Progress on computational architectures, particularly CMOS-based systems, has shown that scaling transistor sizes and optimizing micro-architectures (e.g., variable pipeline depth) can modulate the EDP within physical constraints~\cite{1609253}.
Meanwhile, emerging technologies like reversible computing, spintronics, and neuromorphic systems promise to push EDP closer to fundamental limits by minimizing or circumventing the energy cost of bit erasure~\cite{frank2005introduction}, encoding information in low-energy states~\cite{puebla2020spintronic}, and eliminating costly data movements~\cite{Indiveri_2015}. However, practical challenges—such as noise, interconnect overheads, and thermal management—continue to separate theoretical bounds from achievable performance.



Thermodynamic computing is a leading candidate to accelerate AI workloads, from Bayesian inference to generative AI~\cite{Conte2019Thermodynamic,coles2023thermodynamic_published,Melanson2025,Whitelam2025Generative,Hylton2020,Boyd2025Thermodynamic,Zhang2025}. Moreover, it provides a clear opportunity for exploring energy-time tradeoffs and fundamental bounds on the EDP. Thermodynamic computing algorithms have been developed for matrix inversion and linear systems~\cite{aifer2024_TLA}, matrix exponentials~\cite{PhysRevResearch.7.013147}, Bayesian inference~\cite{aifer2024thermodynamicbayesianinference}, quadratic programming~\cite{bartosik2024thermodynamicalgorithmsquadraticprogramming}, and second-order optimization~\cite{donatella2025scalablethermodynamicsecondorderoptimization}. These algorithms typically involve a key step, which is a quench followed by a relaxation to thermal equilibrium. This protocol is, however, suboptimal in terms of EDP, and recent results  suggest that alternative driving protocols can considerably reduce the energetic cost~\cite{Aurell2011Optimal,abiuso2022thermodynamics,van2023thermodynamic,Whitelam2025Increasing}, 
potentially reaching the limits of the EDP. 


Within this context, in this work we derive fundamental bounds on the EDP of thermodynamic computers and construct driving protocols that can approach them under realistic assumptions. 
We focus on quadratic potentials and equilibrium-based protocols~\cite{abiuso2022thermodynamics}, such as those in Gaussian sampling and linear algebra applications, although our results naturally extend to arbitrary potentials and out-of-equilibrium computations, including recent proposals for non-equilibrium thermodynamic computing~\cite{whitelam2025thermodynamiccomputingequilibrium,Whitelam2025Generative}. 

The paper is structured as follows. First, in \sref{sec:EDDP}, we introduce a key quantity in our work, a natural extension of the EDP that explicitly incorporates the probabilistic nature of thermodynamic computation. 
The resulting energy–delay–deficiency product (EDDP) combines energy, time and reliability into a single metric for thermodynamic algorithms. In \sref{Sec:ThermodynamicComputing}, we characterize the EDDP of a driving protocol and derive a fundamental lower bound on it by leveraging the geometric concept of Wasserstein geodesic~\cite{dechant2019thermodynamicinterpretationwassersteindistance,nakazato2021geometrical}. In \sref{sec:protocols}, we construct realistic driving protocols that can approach such bounds, and apply them to the problem of solving linear systems. 
Finally, we discuss extensions in \sref{sec:Generalizations}, and conclude in \sref{sec:Conclusions}. 

\section{Energy-Delay-Deficiency product (EDDP) in Thermodynamic Computing}
\label{sec:EDDP}


Let us outline the main ideas behind this work. We consider a driven system in contact with a thermal environment at temperature $k_B T = \beta^{-1}$. By suitably controlling the system's potential $V(x,t)$, the dissipative dynamics leads to a state of the system that encodes the solution of a given computation. 
This constitutes a \emph{thermodynamic algorithm} which consists of two basic steps (see also \fref{fig:thermo_alg}):  
\begin{enumerate}
    \item \label{s1} The system's potential $V({x}, t)$ is driven\footnote{We consider any possible time-control protocol, including both continuous protocols but also sudden quenches (followed by equilibration processes where $V(x)$ remains constant).}  from $V_0({x}) := V({x}, 0)$ to $V_1({x}) := V({x}, \tau_D)$ in a fixed amount of time $\tau_D$, and as a consequence the system evolves from the  distribution $\rho_0({x})$ into a target distribution  $\rho_1({x})$ which encodes the solution of the problem. Often, $\rho_1({x})$ will correspond to a Gibbs state, namely
   \begin{align}
       \rho_1({x}) \propto e^{-\beta V_1({x})}~.
    \end{align}
    \item \label{s2} The system is measured and samples are collected during a fixed amount of time $\tau_S$. The outcome of the algorithm is given by
    \begin{align}
        \langle f({x}) \rangle = \frac{1}{\tau_S} \int_{\tau_D }^{\tau_D  +\tau_S}\!\!dt'~ f({x}(t'))~.  
    \end{align}
\end{enumerate}
In the Markovian regime, we expect the error in the estimation in \stepref{s2} to typically decay as $\sim 1/\tau_S$ and, in turn, the irrecoverable energy required to perform \stepref{s1} decays as $\sim 1/\tau_D$ (see below). 
This clearly indicates a trade-off between the accuracy, energy, and total time $\tau := \tau_D+\tau_S$ of the computation. 
In order to account for all relevant resources, we will thus gauge the performance of thermodynamic algorithms via an extension of the well-known energy-delay-product (EDP)~\cite{Rabaey2003}, namely a quantity of the form, 
\begin{align} \label{eq:edqp_prelim}
    \boxed{\text{EDDP} := W_{\text{diss}} \tau (1-\mathcal{Q})}
\end{align}
Here, $\Wir$ is the irrecoverable energy (dissipated work) and $0\leq\mathcal{Q}\leq 1$ is a suitably defined quality factor. 
We will refer to the quantity from \eref{eq:edqp_prelim} as the \emph{energy-delay-deficiency product} (EDDP). 

In what follows, we will quantify this tradeoff for thermodynamic hardware based upon a driven overdamped classical system~\cite{coles2023thermodynamic_published,aifer2024_TLA,aifer2024_TBI}, and argue that similar tradeoffs are valid as well for other physical platforms. Specifically, we will derive fundamental lower bounds to the EDDP and develop simple driving protocols with good-but-not-optimal performance. The key steps in our proofs draw from recent technical results in stochastic thermodynamics~\cite{van2023thermodynamic} which lower bound the energy required to thermodynamically drive between any two states of the system.

\section{Thermodynamic computing in Langevin systems and bounds on EDDP}
\label{Sec:ThermodynamicComputing}
In the core of this work we consider, as base playground for the optimization of thermal computing protocols, physical systems whose dynamics is well described by the so-called (overdamped) Langevin equation~\cite{risken1996fokker}. In terms of the particles' position vector $x$ and energy potential $V(x,t)$, this reads as
\begin{align} \label{eq:overdamped}
    \dot{x} = - \frac{1}{\gamma} \nabla V(x,t) + \sqrt{\frac{2}{\gamma \beta }} \eta(t)\;,
\end{align}
where $\gamma$ is a friction coefficient, and $\eta(t)$ is a white noise vector satisfying $\langle \eta_i(t) \eta_j(t')\rangle = \delta_{ij} \delta(t-t')$.
This equation models systems in which particles' momentum relaxes rapidly, such as colloidal particles in a viscous medium, however notice that several systems have been shown to be well described by the same dynamical equation~\cite{ciliberto2017experiments}, such as particles in noisy optical traps~\cite{gao2017optical}, as well as overdamped RLC circuits~\cite{ciliberto2017experiments,aifer2024_TLA}. 

For a fixed potential $V(x,t)\equiv V(x)$ the probability distribution of the system tends to the thermal state represented by the Gibbs-Boltzmann distribution 
\begin{align}
    \rho^{\rm th}(x) = \frac{1}{Z} e^{-\beta V(x)},
\end{align}
where $Z$ simply corresponds to the normalization of the exponential distribution $Z\equiv\int\!dx~e^{-\beta V(x)}$.
The thermodynamics of this and more general stochastic systems has been well characterized~\cite{sekimoto1998langevin,seifert2012stochastic}. Specifically, the internal energy $U$, work $\delta W$ and heat $\delta Q$ entering the system can be expressed as
\begin{align}
E &= \int \!dx~ \rho(x)V(x)~,\\
 \delta W &=\int \!dx~ \rho(x) \delta V(x)~,\\
   \qquad \delta Q &=\int \!dx~ \delta \rho(x) V(x)~,
\end{align}
where $\delta E=\delta W+\delta Q$ is the first law of thermodynamics.

\stepref{s1} of a thermodynamic algorithm consists in a potential transformation $V_0(x)\rightarrow V_1(x)$ at constant surrounding temperature $k_{\rm B}T=\beta^{-1}$. 
The overall work cost for such transformation is given by
\begin{align}
     W^{(0)\rightarrow (1)}=\Delta F+\Wir^{(0)\rightarrow (1)}\;,
    \label{eq:Wirr_2ndlaw}
\end{align}
where $\Delta F=F^{(1)}-F^{(0)}$ is the variation of the free energy  $F=U-\beta^{-1}S$, with $S=-\int\!dx~\rho(x)\ln\rho(x)$ being the entropy. It follows that $\Wir^{(0) \rightarrow (1)}$ is non-negative according to the second law of thermodynamics. 
\stepref{s2} of a thermodynamic algorithm consist in sampling the position of the particle $x$ which does not modify the energetics of the system. Therefore, no work is performed in this step and control parameters are held fixed.
Notice that the variation in free energy, $\Delta F$, can always be recovered by complementing the protocol with a (sufficiently slow) reversible reset transformation taking $V_1(x)\rightarrow V_0(x)$. In principle, one could also consider more general sequential protocols in which $\{\rho_1^{\rm th},V_1\}$ produced in a previous run is treated as the new initialization for a following run, and so on; the total work required to perform $m$ such thermodynamic algorithms correspond to a sum of contributions for each step, that is
\begin{align}
\sum_{i=1}^{m} W^{(i-1) \rightarrow (i)}&=\sum_{i=1}^{m} \Delta_i F+\sum_{i=1}^{m} \Wir^{(i-1)\rightarrow (i)}\\
&=F^{(m)}-F^{(0)}+\sum_{i=1}^{m} \Wir^{(i-1)},
\end{align}
where $\Delta_i F := F^{(i)} - F^{(i-1)}$. Once again, the global term $F^{(m)}-F^{(0)}$ (which can have any sign) is null in case $V_0\equiv V_m$, or can be made null via a reversible transformation at the end. Finally, in case an infinitely slow reset is forbidden, e.g. due to time constraints, one can still perform the $V_1\rightarrow V_0$ reset in finite time, by reversing the forward protocol. It can be shown (cf. \eref{eq:w_diss_quad}) that in such case $\Wir^{(0)\rightarrow(1)}=\Wir^{(1)\rightarrow(0)}$, and thus the total energy expenditure of the full protocol would be $2\Wir^{(0)\rightarrow(1)}$.

For all these reasons, in the following we will consider the dissipation during \stepref{s1}, namely
\begin{align}
\Wir := W_{\rm diss}^{(0) \rightarrow (1)},
\end{align}
to be the fundamental energy cost associated with a thermodynamic algorithm.

\subsection{Wasserstein geometry and fundamental bounds on EDDP}
We now introduce the main technical tool needed to derive our optimal EDP-like trade-offs. The energy cost $\Wir$ is, in general, protocol-dependent and requires a full solution of the dynamical equations in order to be estimated.
However, one can still obtain valuable insights with the help of geometry: a key result from Ref.~\cite{dechant2019thermodynamicinterpretationwassersteindistance,nakazato2021geometrical} shows that for systems governed by Langevin equation from \eref{eq:overdamped}, the dissipation $\Wir$ can be bounded by the square of the Bures-Wasserstein distance $\mathcal{W}$ between the initial and final distributions, $\rho_0 := \rho_0(x) $ and $\rho_1 := \rho_1(x)$, that is defined as
\begin{align} \label{eq:wass_def}
\mathcal{W}^2(\rho_0, \rho_1) := \inf_{\pi \in \Pi(\rho_0, \rho_1)} \int_{\mathbb{R}^n \times \mathbb{R}^n}\hspace{-20pt} |x - y|^2 d\pi(x, y),
\end{align}
where $\Pi(\rho_0, \rho_1)$ is the set of all probability density functions (so-called transport plans) having $\rho_0$ and $\rho_1$ as marginals~\cite{villani2008optimal}.  
According to Ref.~\cite{dechant2019thermodynamicinterpretationwassersteindistance,nakazato2021geometrical}, isothermal transformations of overdamped Langevin systems universally satisfy
\begin{align} \label{eq:WirWar}
W_{\text{diss}} \geq \frac{1}{\beta\gamma \tau_D} \mathcal{W}^2(\rho_0, \rho_1).
\end{align}
This inequality captures a key physical insight: faster protocols \emph{necessarily} incur larger dissipation. In principle, the dissipation $\Wir$ can even vanish in the limit $\tau_D \to \infty$, reflecting the reversibility of infinitely slow processes. Crucially, \eref{eq:WirWar} can be saturated by protocols in which the instantaneous distribution $\rho_t$ follows the geodesic path in terms of Bures-Wasserstein distance between $\rho_0$ and $\rho_1$~\cite{dechant2019thermodynamicinterpretationwassersteindistance,nakazato2021geometrical}. 

The above connection recasts the thermodynamic task of identifying a minimally dissipative protocol as a geometric problem. 
In particular, \eref{eq:WirWar} immediately provides the ultimate bound on the EDP restricted to \stepref{s1}---namely $\Wir\tau_D$---solely in terms of the temperature, the friction coefficient, and the distance $\mathcal{W}$ covered by the desired computation.
Moreover, consider now the sampling \stepref{s2}. It is safe to assume (for example, if the sampling rate is sufficiently larger than $\gamma^{-1}$) that the result of the computation $\langle f(x)\rangle$ will have, for finite time $\tau_S$, an estimation error  
\begin{align}
    \varepsilon^2:= 
    \left(\frac{1}{\tau_S} \int_{\tau_D }^{\tau_D  + \tau_S}\!\!dt' f({x}(t')) - \int \!dx~f(x)\rho_1(x)\right)^2
    \propto \frac{1}{\tau_S}\;.
\end{align}
For a given total time $\tau=\tau_D+\tau_S$ it then follows that 
\begin{align}
    \frac{\tau_S}{\tau_D+\tau_S}=\frac{\varepsilon^2_{\rm min}}{\varepsilon^2}=:\Q\;,
    \label{eq:error_ratio}
\end{align}
where $\varepsilon^2_{\rm min}$ corresponds to the minimum error achievable in case the entire time $\tau$ of the protocol was spent for \stepref{s2}---sending $\tau_D\rightarrow 0$ is, in principle possible, at the expense of an asymptotically large $\Wir$. Therefore, $0\leq \Q\leq 1$ can be seen to represent the \emph{quality factor} of the computation.

Using \eref{eq:error_ratio}, it follows  that the EDP of \stepref{s1} coincides with the EDDP of the entire algorithm, that is $\Wir\tau_D=\Wir\,\tau\left(1-\Q
\right)$. Combining this observation with the bound in \eref{eq:WirWar}, we  can now state our main result as an EDP-like bound for thermodynamic computing, 
\begin{align}
    \boxed{{\rm EDDP}
    \geq \frac{\mathcal{W}^2(\rho_0, \rho_1)}{\beta\gamma}\;.}
    \label{eq:main_EDP}
\end{align}
That is, as the quality $\Q$ of the computation increases, the dissipated energy or the total time must increase.
In the remainder of the paper, we will estimate $\W$ for relevant classes of protocols and look for practical thermodynamic computations that are close to saturating the fundamental bound~\eq{eq:main_EDP}.

\subsection{Lower bound on \texorpdfstring{$\W$}{W} and quadratic systems}
The Wasserstein distance $\W(\rho_0,\rho_1)$ between two arbitrary distributions, as defined in \eref{eq:wass_def}, is in general nontrivial to estimate, particularly for large dimensions, so it is useful to bound it using only the first two moments. For that let us define
\begin{align}
\mu_i &= \int\! dx~ x \rho_i(x)~,\\ 
\Sigma_i &= \int\! dx~ (x - \mu_i)(x - \mu_i)^T \rho_i(x)~.
\end{align}
Then, following Ref.~\cite{gelbrich1990formula}, one can lower-bound the Wasserstein distance as
\begin{align}
\label{eq:W_low_bou}
\W(\rho_0,\rho_1)^2 &\geq |\mu_0 - \mu_1|^2 +\B\W(\Sigma_0,\Sigma_1)^2~,\\
\B\W(\Sigma_0,\Sigma_1)^2 &:=\Tr[\Sigma_0 + \Sigma_1 - 2 (\Sigma_1^{1/2} \Sigma_0 \Sigma_1^{1/2})^{1/2}]~.\nonumber
\end{align}
where the notation $\B\W(\Sigma_0,\Sigma_1)$ defines the \emph{Bures-Wasserstein distance}~\cite{bhatia2019bures} computed for two covariance matrices $\Sigma_0$ and $\Sigma_1$.

The bound \eref{eq:W_low_bou} is saturated in the case of Gaussian distributions moving along their corresponding Wasserstein geodesics. In the context of thermal equilibrium, Gaussian distributions are exactly those arising from quadratic potentials of the form 
\begin{align}
    V(x) = \frac{1}{2} x^T K x + b^T x
\end{align}
which leads to the equilibrium distribution
\begin{equation}
     \rho^{\rm th}(x) = \frac{e^{-\frac{1}{2}(x-\mu)^T \Sigma^{-1}(x-\mu)}}{\sqrt{(2\pi)^N\det \Sigma}}
\end{equation}
where $\Sigma=(\beta K)^{-1}$, and $\mu=-K^{-1}b$.
Here \(K\) is a positive semi-definite matrix encoding pairwise interactions, and \(b\) is a vector describing linear energy terms. 
The thermodynamics of quadratic overdamped systems has been characterized in Ref.~\cite{abiuso2022thermodynamics}.

The corresponding entropy production can be then expressed from the integral over time of the euclidean metric for $\mu$ and the $\B\W$ metric for $\Sigma$, i.e.
\begin{align}
    \label{eq:w_diss_quad}
    \beta\Wir =&\int_0^\tau\!dt~ |\dot{\mu}(t)|^2 \\+&  \frac{1}{2}\!\int_{0}^{\tau}\!\!dt\Tr[\int_{0}^{\infty}\!\!\!d\nu~e^{-\nu \Sigma(t)} \dot{\Sigma}(t) e^{-\nu \Sigma(t)} \dot{\Sigma}(t) ].
    \nonumber
\end{align}
In our applications, we will in general assume $\mu$ constant as it only provides an irrelevant shift of the distribution. Moreover one can often assume \(\Sigma_0 \propto \mathbb{1}\), which simplifies the analysis. In particular, when \(\Sigma_0\) and \(\Sigma_1\) commute, i.e., \([\Sigma_0, \Sigma_1] = 0\), the squared $\B\W$ distance decomposes into a sum over the eigenvalues \(\sigma_{i,\alpha}\) of \(\Sigma_i\), yielding the closed-form formula $\B\W(\Sigma_0,\Sigma_1)^2 = \sum_\alpha (\sqrt{\sigma_{0,\alpha}} - \sqrt{\sigma_{1,\alpha}})^2$.

\subsection{General optimization principles: symmetries and slow-driving}
\label{sec:sym_and_slow}

The expressions for the dissipated $\Wir$~\eq{eq:w_diss_quad} and its achievable minimum $\B\W$~\eq{eq:W_low_bou} will thus be central in the design of optimal and close-to-optimal protocols for (Gaussian) thermodynamic computing. Before dwelving into explicit protocols, it is  worth discussing a few considerations.

First, as $\beta^{-1}$ naturally sets the energy scale of the problem, it is easy to notice that the transformation $\beta^{-1}\rightarrow\lambda\beta^{-1}$ leads to
$ \Sigma(t)\rightarrow\lambda\Sigma(t)$
which, upon substitution into \eref{eq:w_diss_quad}, yields immediately
$\Wir\rightarrow \lambda\Wir$.
Thus, $\Wir$ is \emph{linear} in the temperature as expected.

Secondly, we can note that a scale invariance appears  of the dissipation naturally emerges from \eref{eq:w_diss_quad}: for a given finite-time protocol $K(t)$ the dissipation remains constant for the transformation $K(t) \rightarrow K'(t)=\lambda K(\lambda t)$ (with $\tau_D \rightarrow \tau_D/\lambda$ and $\lambda>0$), i.e. scaling it in magnitude by a factor $\lambda$ and accelerating it by the same factor leaves the dissipation unchanged.
In fact, one can see that given a solution $\Sigma(t)$ of the dynamics associated to the driving $K(t)$, the rescaled $\Sigma'(t)=\lambda^{-1}\,\Sigma\bigl(\lambda t\bigr)$ corresponds to a solution for $K'(t)$.
Substituting this into \eref{eq:w_diss_quad}, the same transformation provides $\Wir\rightarrow\Wir$, which is therefore
invariant under this time-strength rescaling (more details in \aref{app:symmetries}).  This symmetry implies that amplifying the protocol’s strength by $\lambda$ while compressing its duration to $\tau/\lambda$ yields an equivalent thermodynamic outcome. This scaling symmetry defines an equivalence class in the family of protocols with the same EDDP. 
A direct consequence is that in principle one should use the highest value of $\lambda$ allowed by the physical hardware at hand. Indeed, any computation encoded in $\rho_1(x)$ obtained from $\rho_0(x)$ is equivalently encoded in $\rho_1'(x)\propto\rho_1(\sqrt{\lambda} x)$ obtained from $\rho_0'(x)\propto\rho_0(\sqrt{\lambda} x)$ in a protocol pertaining to the same equivalence class. However any physical device will entail a maximum value of $\lambda$, due to both a finite achievable strength of the energy potential $K$, and a finite resolution when sampling from a distribution with spread $\Sigma$.

Finally, let us mention that we will often consider, in our analysis, the regime in which $\tau_D$ and $\tau_S$ are much larger than the thermalization rate $\gamma^{-1}$. In fact, a large $\tau_S$ is needed to yield a small computation error $\varepsilon^2$, whereas a large $\tau_D$ enables the possibility of small dissipation according to \eref{eq:WirWar}. In terms of dynamical driving this is often referred to as \emph{slow-driving} regime~\cite{sivak2012thermodynamic,cavina2017slow,abiuso2020geometric}, which assumes the rate of change of $V(x,t)$ in time is sufficiently slow to have the state
$\rho(x,t)$ always close to the thermal $\rho^{\rm th}(x)$ corresponding to $V(x,t)$. More specifically one can expand
\begin{align}\label{eq:slow_approx}
    \Sigma(t)=\beta^{-1} K^{-1}(t)+\mathcal{O}\left(\frac{1}{\tau_D}\right)
\end{align}
and substitute it in \eref{eq:w_diss_quad} to express $\Wir$ up to $\mathcal{O}\left(\tau_D^{-2}\right)$ corrections~\cite{abiuso2022thermodynamics}. That is, in the slow-driving regime, we do not need to solve the dynamical equations for $\rho(x,t)$ and $\Sigma(t)$, rather we can express the thermodynamic performance directly in terms of the driving $K(t)$.

\section{EDDP of Protocols with limited control and available information} 
\label{sec:protocols}
Up to this point, our analysis has focused on the controlled time evolution of the system’s distribution $\rho(x,t)$. In practice, however, one cannot directly manipulate the system’s state, but only its energy potential $V(x,t)$. In this section we will focus on characterizing and optimizing with different approaches and classes of protocols the EDP incurred during \stepref{s1} of the algorithm---and thus the EDDP of the algorithm as a whole---for Gaussian thermodynamic computing.

\subsection{Quenches}
Let us start by considering protocols that require the least amount of control and knowledge: quenches. These consist in an immediate and instantaneous transformation of the potential $V(x,0) = V_0(x)$ to $V(x,0^+) = V_1(x)$ with a fixed waiting time to let the system be sufficiently close to its equilibrium state once we start \stepref{s2}. The dissipated energy of these quenches is fixed by the boundary conditions of the protocol and does not depend on the waiting time for equilibration since we are assuming that the final state is close to equilibrium (cf. \aref{app:protocols})
\begin{equation}
    W_{\text{diss}}^{Q} = \frac{1}{2}k_BT \left( \Tr [K_1K_0^{-1} - \mathbb{1}] +  \ln \!\left[\frac{\det K_0}{\det K_1}\right] \right),
\end{equation}
which leads an energy-delay-deficiency product that scales linearly with $\tau_D$: $\text{EDDP}^Q = W_{\text{diss}}^Q\tau_D$.

\begin{figure*}[ht]
    \centering
    \includegraphics[width=\textwidth]{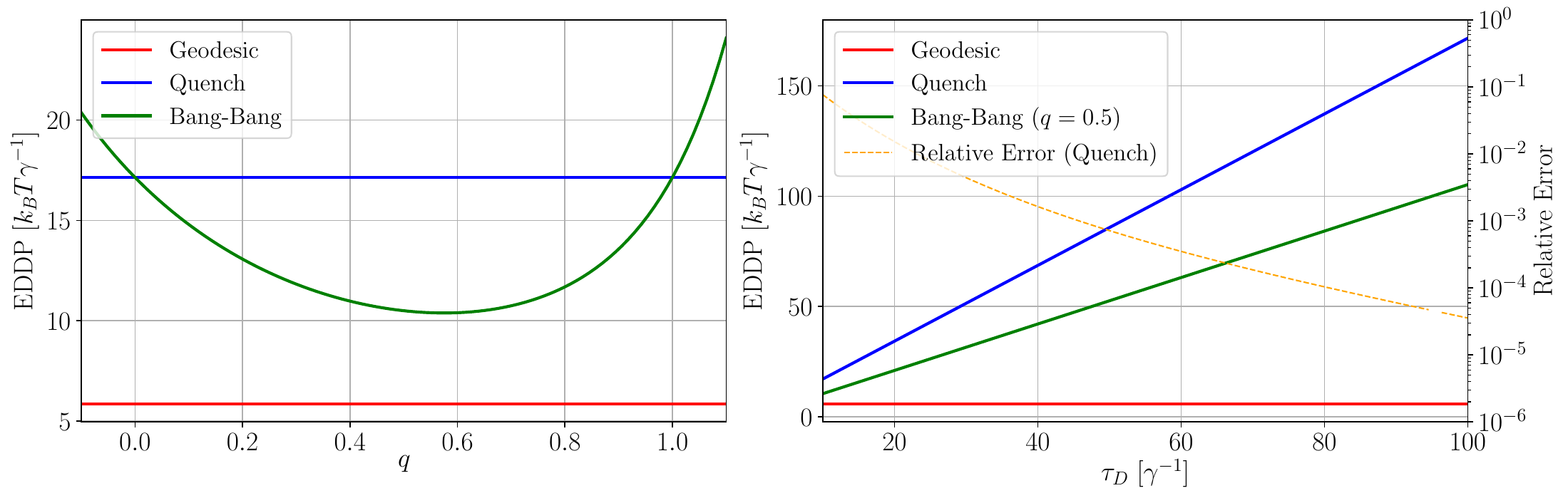}
    \vspace*{-20pt}
    \caption{Average EDDP for 500 samples of $K_1$ taken from a Wishart distribution of $10\times10$ matrices with expected value $K_0 = \mathbb{1}$ and $20$ degrees of freedom. (Left) Comparison of the EDDP for geodesic, quench, and Bang-Bang protocols as a function of the interpolation parameter $q$ for a driving time $\tau_D = 10\gamma^{-1}$. Qualitatively, the shape of the Bang-Bang EDDP curve remains the same as one increases the driving time. (Right) Comparison of the EDDP for geodesic, quench, and Bang-Bang with $q=0.5$ protocols as a function of the driving time $\tau_D$. And relative error of the quench protocol ($\sqrt{\Tr[(\Sigma(\tau_D)-K_1^{-1})^2]/\Tr[K_1^{-2}]}$) as a function of the driving time. The relative error of the Bang-Bang protocol is slightly larger (and dependent on $q$) but is of the same order.}
    \vspace*{-15pt}
    \label{fig:bang-bang}
\end{figure*}
\subsection{Geodesics}
However, the optimal achievable EDDP given by the Bures-Wasserstein distance does not scale with $\tau_D$: $\min_{(K(t),\Sigma(t))} \text{EDDP} = \mathcal{BW}(\Sigma_0,\Sigma_1)/(\beta \gamma)$. From this minimal EDDP we can see how the constant $\beta\gamma$ sets the scale of the problem, which is a consequence of the symmetries discussed in the previous section. Since a rescaling of these constants does not affect the minimization problem, we will be setting $\beta=1$ and $\gamma=1$ for the rest of this section, to lighten the notation.
The minimal EDDP is realized by the geodesic\footnote{It is called a geodesic since the length of this trajectory minimizes the Bures-Wasserstein distance.} state trajectory 
\begin{equation}
\Sigma_{\text{geo}}(t) = \left(\frac{t}{\tau_D}\sqrt{\Sigma_1} + \frac{\tau_D-t}{\tau_D}\sqrt{\Sigma_0}\right)^2~.
\end{equation}
This trajectory can be obtained with the following control (cf. \aref{app:protocols})
\begin{align}\label{eq:Kgeo}
    K_{\rm geo}(t) &= \left[\mathbb{1} - \frac{\chi_{(0,\tau_D)}(t)}{\tau_D}\Delta_{\rm geo}(t)\right]\Sigma_{\text{geo}}(t)^{-1}~,\\
    \Delta_{\rm geo}(t)&:= \frac{t}{\tau_D}\Sigma_1 + \frac{\tau_D-2t}{\tau_D}\sqrt{\Sigma_1}-\frac{\tau_D-t}{\tau_D}\mathbb{1}
\end{align}
with $\chi_{(0,\tau_D)}(t)$ the indicator function of the open interval $(0,\tau_D)$.
Furthermore, this protocol has another advantage compared to the naive quench: it reaches exactly the desired state with covariance matrix $\Sigma_1$ at $\tau_D$. The geodesic protocol achieves this property by ``overshooting'' the final control $K_1$ in such a way that at $t=\tau_D$ the state can reach $\Sigma_1$ (instead of converging asymptotically to it) and then making a quench ``back to'' $K_1$.

However, both the continuous part and the quenches of this geodesic control trajectory require knowledge of $\Sigma_1$. In general, the objective of the algorithm is precisely to extract information from the distribution $\rho_1$. Therefore, we will have to assume that we do not have prior knowledge of this distribution---thus of $\Sigma_1$---and cannot implement the geodesic protocol in \stepref{s1}. 

\begin{figure*}[ht]
    \centering
    \includegraphics[width=\textwidth]{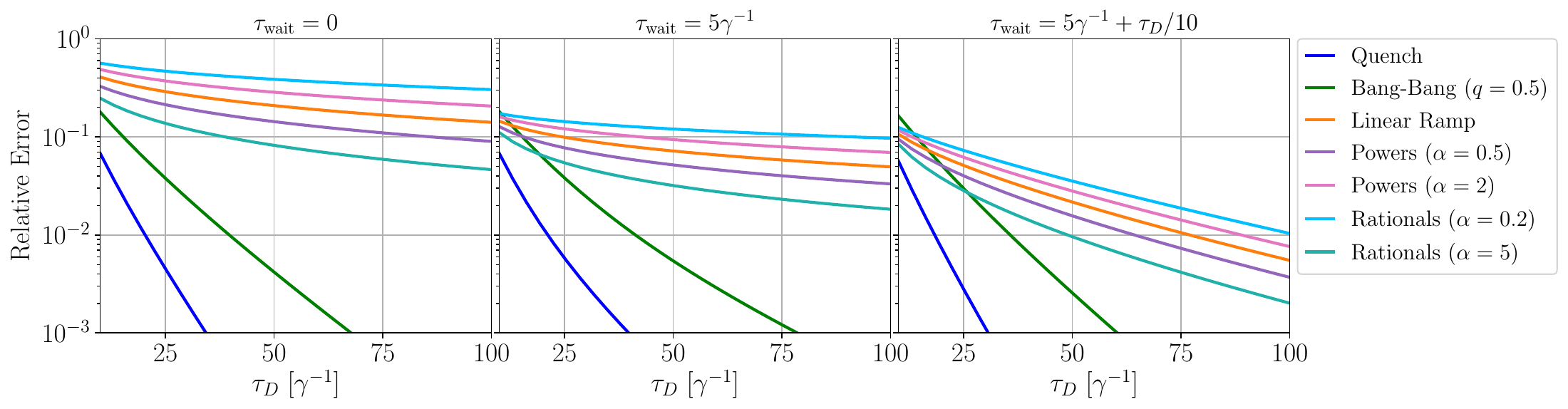}
    \vspace*{-20pt}
    \caption{Average Relative error ($\sqrt{\Tr[(\Sigma(\tau_D)-K_1^{-1})^2]/\Tr[K_1^{-2}]}$) for 100 samples of $K_1$ taken from a Wishart distribution of $10\times10$ matrices with expected value $K_0 = \mathbb{1}$ and $20$ degrees of freedom. The relative error is shown for the different protocols as a function of $\tau_D$ ($\tau_D \in [10\gamma^{-1},100\gamma^{-1}]$) for multiple values of $\tau_{\rm wait}$: (Left) no relaxation time, (Middle) fixed waiting time, (Right) adaptive waiting time.}
    \vspace*{-15pt}
    \label{fig:error_drive}
\end{figure*}

\subsection{Bang-bang protocols}
In what follows we will study how much we can improve on the quenches protocol and get close to the minimal EDDP with protocols that only require interpolations of the matrices $K_0$ and $K_1$. In the interest of outlining what we have access to and what we don't, we will evaluate the EDDP of these protocols in the case study of a specific algorithm: matrix inversion. In this case we are given as input the matrix $K_1$ and desire to compute its inverse. Since the covariance matrix of the thermal state is given by $\Sigma_{th} = K^{-1}$, to compute the inverse in \stepref{s1} we drive the control from $K_0$ to $K_1$, and then we reconstruct the covariance matrix by sampling the position of the colloidal particles during \stepref{s2}. Using the scale symmetries of the cost (outlined in \sref{sec:sym_and_slow} and detailed in \aref{app:symmetries}), we notice that we can take, up to a scalar renormalization of the entire problem, $K_0 = \mathbb{1}$. Moreover, partial knowledge on the set of matrices from which $K_1$ is sampled, can also be used to optimize the initialization scale $K_0 = \mathbb{1}\rightarrow K_0 = \lambda \mathbb{1}$, see \aref{app:scale_optimization}.

One of the simplest improvements one can obtain from the quenches is to split these in two steps, as it has been shown that the energy-optimal protocols for rapidly driven systems are ``bang-bang'' protocols~\cite{rolandi2023fast}. The procedure consists in a first quench at $t=0$ from the initial control $K_0$ to an intermediary control $K_*$, followed by letting the system thermalize for $\tau_D/2$, then a second quench to the final control $K_1$ and letting the system relax for the remaining time $\tau_D/2$. We obtain that the dissipated work is (cf. \aref{app:protocols})
\begin{equation}
    W_{\text{diss}}^{\text{Bang-Bang}} = W_{\text{diss}}^{Q} - \frac{1}{2}\Tr [(K_1-K_*)(K_0^{-1} - \Sigma_*)],
\end{equation}
where $\Sigma_* = e^{-K_*\tau_D/2}\Sigma_0 e^{-K_*\tau_D/2} + (\mathbb{1}-e^{-K_*\tau_D})K_*^{-1}$ is the covariance of the state at $t=\tau_D/2$. Since we restrict ourselves to only have access to interpolations of $K_0$ and $K_1$ we can naturally write the intermediary control as $K_* = q K_1 + (1-q)K_0$, where $q$ is the interpolation parameter. To evaluate the performance of this protocol we compute its corresponding EDDP for instances of $K_1$ sampled from a Wishart distribution with mean $K_0 = \mathbb{1}$. In \fref{fig:bang-bang} we showcase the mean EDDP for the Bang-Bang protocol, the quench protocol, and the geodesic protocol as a function of the interpolation parameter $q$ and as a function of the driving time $\tau_D$. We only consider driving times larger than $10\gamma^{-1}$ since we need the $\Sigma(\tau_D)$ to be relatively close to the desired thermal state $K^{-1}$ for the quench and Bang-Bang protocols (for the geodesic protocol the error is null). We quantify the relative error with the Hilbert-Schmidt norm $\sqrt{\Tr[(\Sigma(\tau_D)-K_1^{-1})^2]/\Tr[K_1^{-2}]}$. Overall, we can observe that the Bang-Bang protocol yields a significant improvement when the interpolation parameter is around $q\approx 0.5$. However, the scaling of the EDDP with respect to $\tau_D$ remains linear. 

\begin{figure}[ht]
    \centering
    \includegraphics[width=.9\linewidth]{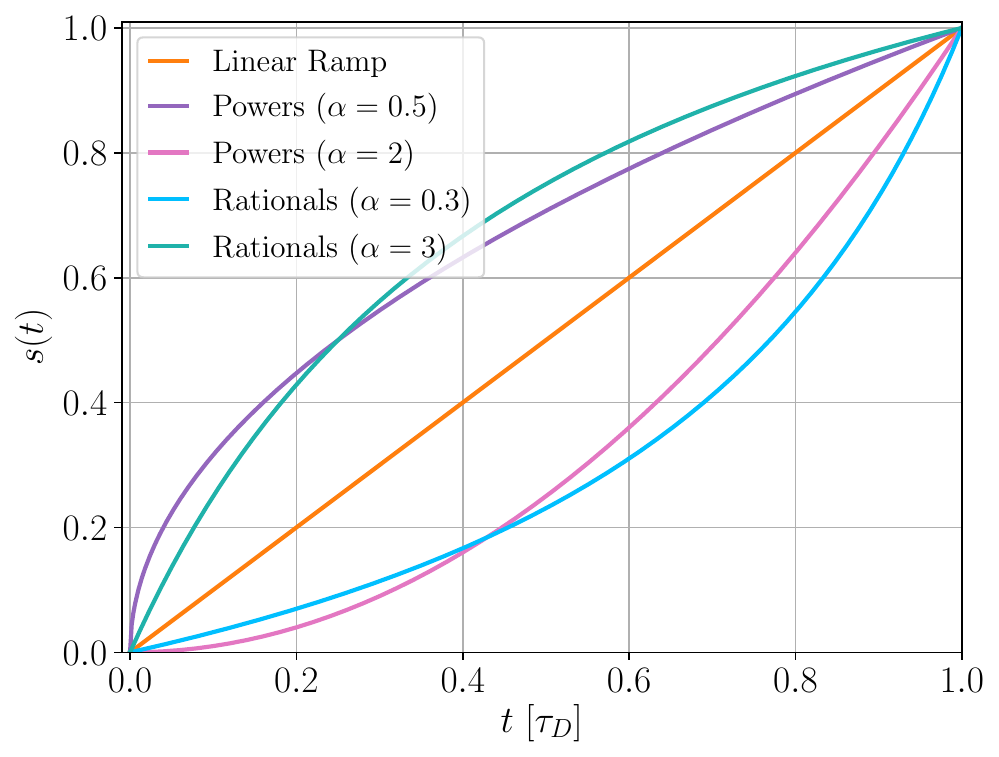}
    \vspace*{-10pt}
    \caption{Examples of continuous driving functions for the Powers and Rationals families.}
    \vspace*{-10pt}
    \label{fig:driving_ex}
\end{figure}
\begin{figure*}[ht]
    \centering
    \includegraphics[width=\textwidth]{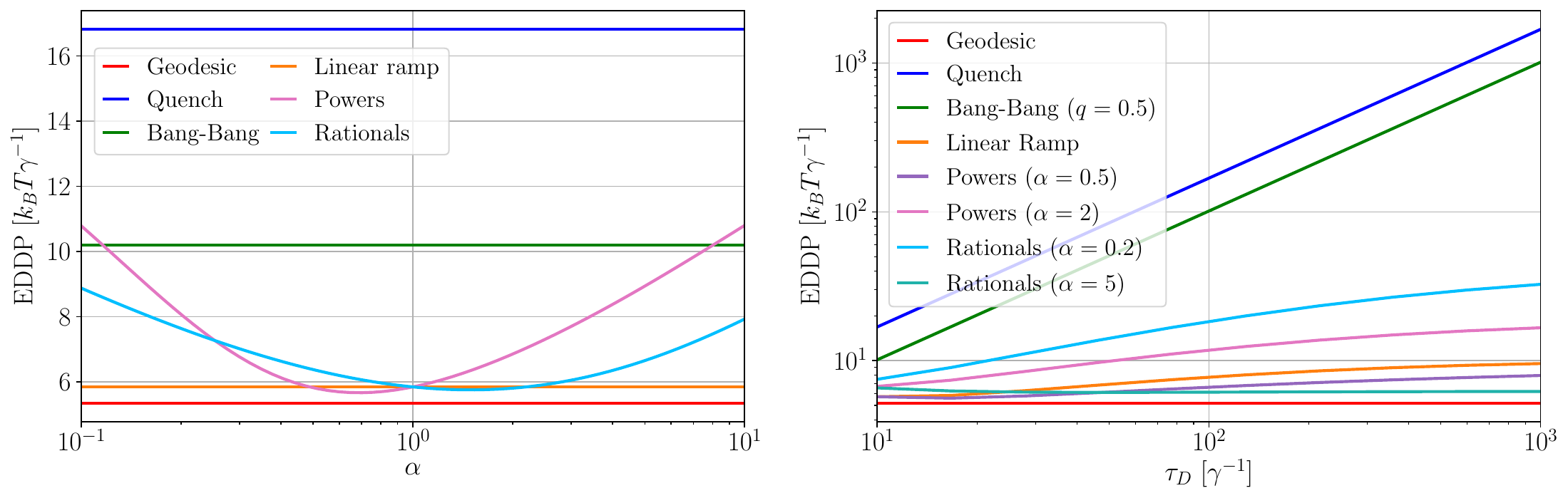}
    \vspace*{-20pt}
    \caption{Average EDDP for 100 samples of $K_1$ taken from a Wishart distribution of $10\times10$ matrices with expected value $K_0 = \mathbb{1}$ and $20$ degrees of freedom. (Left) EDDP of the Powers, and Rationals protocol families as a function of the free parameter $\alpha$ with $\tau_D = 10\gamma^{-1}$ and $\tau_{\rm wait} = 5\gamma^{-1}$. These are compared to the EDDP of the geodesic, quench, Bang-Bang ($q=0.5$), and linear ramp protocols. (Right) Comparison of the EDDP for a collection of protocols as a function $\tau_D$.}
    \vspace*{-15pt}
    \label{fig:driven_protocols}
\end{figure*}
\subsection{Continuous protocols}
We can make a more sophisticated improvement by considering protocols that drive the control continuously from $K_0$ to $K_1$. Since we restrict ourselves to interpolations of the boundary conditions, we can write such protocols as $K(t) = s(t) K_1 + (1-s(t))K_0$, where $s(t)$ is a bijective non-decreasing\footnote{Given the geometric nature of the problem, it is immediate that non-bijective maps do not need to be considered in this problem since for any non-bijective map one can trivially construct a bijective map that will have a smaller dissipation.} map from $[0,\tau_D]$ to $[0,1]$. Here, we will consider two families of functions for $s(t)$:
\begin{align}
    s_{\rm Powers}(t) &= \frac{t^\alpha}{\tau_D^\alpha}~,\\
    s_{\rm Rationals}(t) &= \frac{\alpha t}{\tau_D + (\alpha-1)t}~,
\end{align}
for $\alpha>0$ a free parameter. At $\alpha = 1$ these functions coincide with a linear ramp: $K(t) = \frac{t}{\tau_D}K_1 + \frac{\tau_D-t}{\tau_D}K_0$. We present a visualization of a few examples of protocols in these families in \fref{fig:driving_ex}. It is worth noting that none of these will universally approximate well the geodesics. Indeed, if we consider the concavity of the optimal control $K_{\rm geo}(t)$ in \eref{eq:Kgeo} for scalars (i.e. by taking the trivial case of $1\times1$ matrices) we can see that it is always convex in the bulk. While for these families of interpolated protocols, the convexity switches to concavity (and vice-versa) when $K_0-K_1$ changes sign.

As it was alluded to in the discussion above, unless the driving time $\tau_D$ is large compared to the thermalization timescale $\gamma^{-1}$ the state of the system at the end of \stepref{s1} will not be close enough to the thermal distribution to start the sampling in \stepref{s2}. Therefore, before analyzing the EDDP of the continuous protocols, we start by analyzing the relative error they yield at the end of \stepref{s1}. By \eref{eq:slow_approx} the relative error of $\Sigma(\tau_D)$ scales as $(\gamma\tau_D)^{-1}$ for the continuously driven protocols. However, it is worth noting that once the control is fixed at $K_1$ the state converges exponentially fast to the thermal distribution. Therefore for the quench and bang-bang protocols the error scales as $e^{-\gamma\tau_D}$. By the same reasoning, any error that is present at the start of \stepref{s2} will also be suppressed exponentially fast during the sampling since the state keeps thermalizing during the sampling.

In order to minimize the error, we can also exploit this exponential relaxation for continuously driven protocols by dedicating some of the driving time solely to thermalization. More precisely, we can subdivide the \stepref{s1} of continuously driven protocols in two parts: a. driving the system continuously from $K_0$ (at $t=0$) to $K_1$ (at $t=\tau_D-\tau_{\rm wait}$), b. let the system thermalize for a duration $\tau_{\rm wait}$ (from $t=\tau_D-\tau_{\rm wait}$ to $t=\tau_D$). In \fref{fig:error_drive} we show the relative error of $\Sigma(\tau_D)$ as a function of $\tau_D$ for different amounts of waiting time. In the left panel, we can see that since the error of the continuously driven protocols decays slowly (compared to the quench and bang-bang protocols), it takes significantly long protocols to obtain a reasonable relative error when there is no waiting time. By adding a constant waiting time (middle panel) we improve the situation by a constant factor, however the scaling remains the same. The scaling can be improved by taking a $\tau_{\rm wait}$ that scales with $\tau_D$ (right panel), which will allow for an exponential decay of the error as in the quench and bang-bang protocols. In what follows, we will restrict ourselves to cases with a constant waiting time as it is sufficient for our analysis to have a reasonably small relative error at the end of \stepref{s1}.

In \fref{fig:driven_protocols} we study how the EDDP of these continuous protocols changes depending on the free parameter $\alpha$ and compare it to the performance of the geodesic, quench, and Bang-Bang protocols. We can see that for relatively fast protocols ($\tau_D = 10$ in the left panel) the continuous protocols perform much better than the quenches and Bang-Bang as they are quite close to the geodesic. Furthermore the linear ramp, which is also the most naive guess for a continuous protocol, is close to optimal for both families of protocols. In terms of scaling with respect to $\tau_D$ the continuous protocols perform significantly better than the quench and Bang-Bang protocols. From the right panel we can clearly see that the EDDP of the continuous protocols scales sub-linearly with $\tau_D$. However, we can further prove that it converges to a constant in the slow driving limit: by inserting \eref{eq:slow_approx} into \eref{eq:w_diss_quad} and using the fact that $[K(t),K(t')] = 0$ along these protocols (since we chose $K_0=\mathbb{1}$) we find
\begin{equation}\label{eq:EDP_slow}
    {\rm EDDP} = \frac{\tau_D}{4}\int_0^{\tau_D}\!\!dt~\Tr\!\left[\frac{\dot K(t)^2}{K(t)^3}\right] + \mathcal O\!\left(\frac{1}{\tau_D}\right)~.
\end{equation}
By construction, the interpolation is such that $\dot s(t)\propto \tau_D^{-1}$, therefore we can show that the leading order of \eref{eq:EDP_slow} is independent of $\tau_D$. Furthermore, since the sampling time $\tau_S$ in \stepref{s2} needs to be much larger $\gamma^{-1}$, we are free to assume the same for $\tau_D$ and enter the slow-driving regime (while still being able to keep it small compared to $\tau_S$ and keeping a large quality factor $Q$). In this scenario, we can ignore the waiting time since in the slow driving regime the state of the system remains close to the thermal distribution throughout the protocol.

\subsection{Slow driving protocols}
\begin{figure*}[ht]
    \centering
    \includegraphics[width=\textwidth]{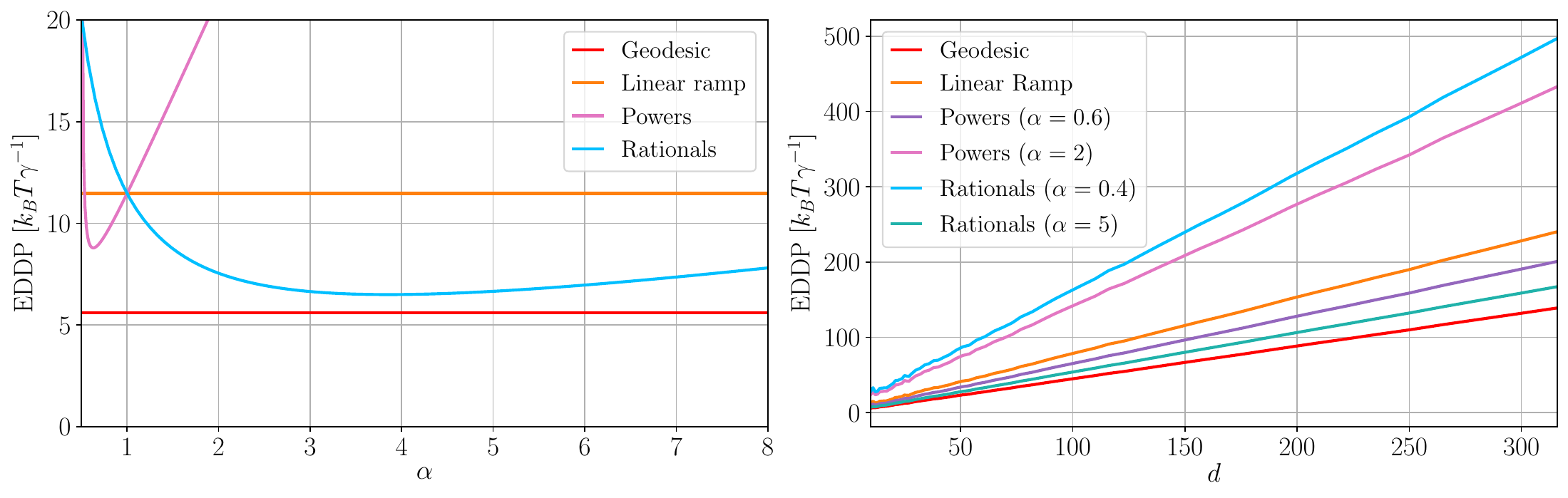}
    \vspace*{-20pt}
    \caption{Average EDDP in the slow-driving regime for 200 samples of $K_1$ taken from a Wishart distribution of $d\times d$ matrices with expected value $K_0 = \mathbb{1}$ and $2d$ degrees of freedom. (Left) EDDP of the Powers, and Rationals protocol families as a function of the free parameter $\alpha$ with $d = 10$. These are compared to the EDDP of the geodesic and linear ramp protocols. (Right) Comparison of the EDDP for a collection of protocols as a function of the matrix dimension $d$.}
    \vspace*{-5pt}
    \label{fig:slow_protocols}
\end{figure*}
\begin{figure}[ht]
    \centering
    \includegraphics[width=\linewidth]{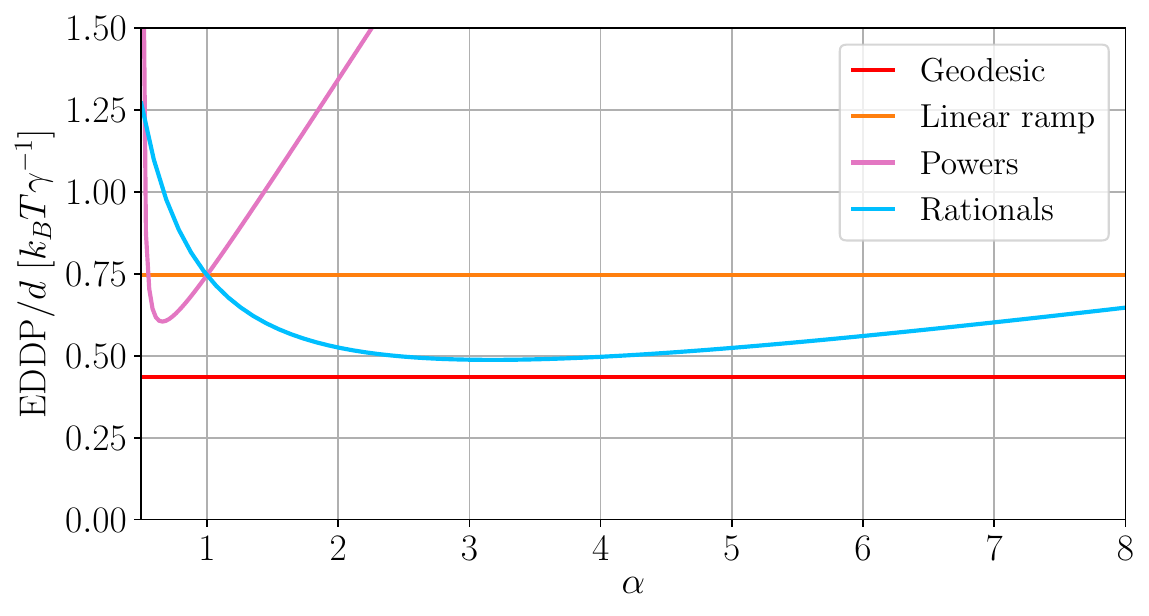}
    \vspace*{-15pt}
    \caption{Average scaling of the slow-driving EDDP as a function of the matrix dimension $d$ (i.e. EDDP$/d$) for 150 samples of $K_1$ taken from a Wishart distribution of $d\times d$ matrices with expected value $K_0 = \mathbb{1}$ and $2d$ degrees of freedom. We compare the EDDP scaling for the Powers, and Rationals protocol families as a function of the free parameter $\alpha$. These are compared to the EDDP scaling of the geodesic and linear ramp protocols.}
    \vspace*{-5pt}
    \label{fig:slow_scaling}
\end{figure}

Since for the quench and Bang-Bang protocols EDDP scales with $\tau_D$, it diverges in the slow-driving regime. Therefore, we cannot compare it to the EDDP of the continuous protocols. In \fref{fig:slow_protocols} we study the EDDP in the slow-driving regime for the different continuous protocols as a function of the free parameter $\alpha$ and the dimension of the matrices. Generally, we can notice that most continuous protocols we consider perform relatively close to the optimal EDDP given by the geodesic. For example, the Rationals protocol with $\alpha\approx 4$ is significantly close to the theoretical optimum. It is also worth noting that we can trade-off EDDP for simplicity by implementing the Linear Ramp protocol, which performs with approximately double the EDDP of the geodesic. We also study how the EDDP depends on the matrix dimension $d$, observing a linear scaling that emerges as the dimension of the matrix increases, and that it is highly dependent on the specific protocol. Indeed, this can be proven from \eref{eq:EDP_slow}: since the protocol satisfies $[K(t),K(t')] = 0$ we can rewrite it as a sum over the eigenvalues $k_i(t)$ of $K(t)$
\begin{equation}
    {\rm EDDP} = \frac{\tau_D}{4}\sum_{i=1}^d\int_0^{\tau_D}\!\!dt\frac{\dot k_i(t)^2}{k_i(t)^3} + \mathcal O\!\left(\frac{1}{\tau_D}\right)~.
\end{equation}
Since the eigenvalues are identically distributed we can show that the EDDP scales linearly with the dimension as $d$ becomes large (cf. \aref{app:protocols}). We can therefore analyze the scaling of the slow-driving EDDP as a function of the free parameter $\alpha$, which we show in \fref{fig:slow_scaling}. What we can observe is very similar to what we have found already for $d=10$, however, the protocol that gets closest to the theoretical optimum becomes the rational protocol with $\alpha\approx3.2$.


\section{EDDP bounds for arbitrary time-dependent potentials and thermodynamic neural networks}
\label{sec:Generalizations}
We now summarize how our geometric approach for quantifying the EDDP trade-off extends beyond the overdamped Brownian particle in a quadratic potential. The same geometric approach, namely measuring entropy production via a transport distance, applies to arbitrary potentials, to underdamped (phase-space) Langevin dynamics, and even to discrete Markov jump processes, albeit with important caveats on computability and tightness. Below we discuss these extensions in greater detail.

In this regime, we retain the overdamped Langevin model from \eref{eq:overdamped}, but now allow for an arbitrary potential $V(x,t)$. The resulting bound on entropy production keeps the same structure, namely, 
\begin{align}
    \label{eq:time-dependent}
   W_{\mathrm{diss}}\;\ge\;\frac{1}{\beta\gamma\tau_D}\,\mathcal{W}^2(\rho_0,\rho_1), 
\end{align}
In practice, the transport distance $\mathcal{W}$ can be difficult to compute for general (non-Gaussian) $\rho_{0,1}$. Importantly, the bound on the Bures-Wasserstein distance from \eref{eq:W_low_bou} is valid for arbitrary potentials and uses only the first two moments of the distribution, namely
\begin{multline}
    \mathcal{W}^2(\rho_0,\rho_1)\;\ge\; \|\mu_0-\mu_1\|^2+\Tr\Sigma_0+\Tr\Sigma_1
\\
-2\Tr\!\big[(\Sigma_0^{1/2}\Sigma_1\Sigma_0^{1/2})^{1/2}\big],
\end{multline}
with $\mu_{0,1}$ and $\Sigma_{0,1}$ being the means and covariances of $\rho_{0,1}$. This allows us to bound the EDDP product even when the endpoints of the protocol are not simply Gaussian states, which occurs when the driving protocol is non-quadratic. 

Indeed, not all thermodynamic algorithms require access to a Gibbs state. Recent work in this direction, see in particular Refs.~\cite{whitelam2025thermodynamiccomputingequilibrium,Whitelam2025Generative,tsormpatzoglou2025thermodynamics}, show an interesting way to use Langevin dynamics for computation using arbitrary potentials, with sampling occurring at a chosen observational time, i.e. without requiring full equilibration. We discuss this approach briefly below in the context of our results. 

Thermodynamic neural networks provide an example of a useful Langevin-based computation that does not require full equilibration and does not rely on quadratic potentials~\cite{whitelam2025thermodynamiccomputingequilibrium}. A single thermodynamic neuron is a scalar degree of freedom confined by a higher-order potential
given by
\begin{align}
    U_J(x, I) = - I x + J_2 x^2 + J_3 x^3 + J_4 x^4,
\end{align}
where $J = (J_2, J_3, J_4)$ specifies the couplings and $I$ denotes an input (bias) signal. Networks are formed by coupling such neurons via bilinear terms so that the total potential of the device reads
\begin{align}
    V(x, I) = \sum_i U_J(x_i, b_i) + \sum_{<i,j>} J_{ij} x_i x_j + \sum_{i, j} W_{ij} I_i x_j.
\end{align}
The time evolution of such a device is governed by the overdamped Langevin dynamics from \eref{eq:overdamped}. In this paradigm one trains the network using e.g. gradient-based method on a digital device, so that the thermodynamic device produces a desired function of the input at a prescribed measurement time $\tau_{M}$. This output is obtained similar to the protocols discussed in this work, that is, sampling the network at the measurement time $\tau_M$ and then averaging.

In the context of our results, this class  of finite-time computations also obeys a similar type of EDDP as in \eref{eq:main_EDP} that trades sampling time, dissipation and quality of computation. The only difference is that now the lower bound on dissipation from \eref{eq:WirWar} must be expressed in terms of the distance between the actual distributions that characterize the device at the initial and final point of the driving protocol. It would be interesting to see if the driving protocols discussed in \sref{sec:protocols} (e.g. the Rationals protocol) are also close to the optimal protocol specified by the geodesic. 

Finally, we note that the considered bounds can also be extended to underdamped dynamics and discrete systems. In the former case, the over-damped bounds  from \eref{eq:time-dependent} can be applied to the position marginals of the phase-space densities. This yields valid lower bounds on dissipation, but momentum degrees of freedom typically prevent the corresponding bound from being tight. 
Similar geometric bounds for discrete systems have been discussed in Ref.~\cite{van2021geometrical}. Importantly, they admit useful lower bounds by operational metrics (e.g. total variation or relative entropy bounds), which can yield model-independent estimates of minimal dissipation for discrete thermodynamic algorithms.

\section{Conclusion and Outlook}
\label{sec:Conclusions}

In this work, we characterized the energy–delay product (EDP) of thermodynamic algorithms based on Langevin dynamics, in which dissipative evolution naturally drives the system toward the solution of the computation. We first derived general bounds on the energy–delay–deficiency product (EDDP)—a generalization of EDP that incorporates the stochastic nature of thermodynamic computing—by exploiting geometric bounds on dissipation~\cite{dechant2019thermodynamicinterpretationwassersteindistance,nakazato2021geometrical}. Although these bounds can, in principle, be saturated by physical evolutions, the corresponding optimal driving protocols require knowledge of the final equilibrium state, i.e., the solution itself. This motivated the development of quasi-optimal protocols using only partial information and limited control, which we demonstrated in the context of matrix inversion. The derived protocols do not require prior information and approach the fundamental thermodynamic limits.

While our analysis focused on linear inversion problems for overdamped quadratic systems, the EDDP bounds extend to general potentials and to underdamped dynamics---although they are tight only for quadratic Hamiltonians and Gaussian states~\cite{abiuso2022thermodynamics}. In particular, they apply directly to the recently proposed framework of thermodynamic networks in Ref.~\cite{whitelam2025thermodynamiccomputingequilibrium}.

Looking ahead, it is natural to explore similar trade-offs in other models of thermodynamic computation beyond Langevin systems, e.g., Refs.~\cite{LipkaBartosik2024,vovchenko2025thermodynamic}, and in physics-based computing more broadly. Related energetic constraints have already been observed in minimal thermodynamic logic gates~\cite{wolpert2020thermodynamics,freitas2021stochastic,Gao2021} and in energy-based machine learning models~\cite{hnybida2025minimal}. The generalized EDDP provides a natural metric to capture such trade-offs, and we expect that unifying principles—analogous in spirit to thermodynamic uncertainty relations—may ultimately yield general bounds on EDDP across a wide class of physical computing architectures. Such results could offer concrete guidelines for designing computing hardware and algorithms that operate closer to the fundamental energetic limits.

\acknowledgments
We warmly thank Nicolas Brunner, Sadra Boreiri, Jake Xuereb, Sam Duffield, and Denis Melanson for fruitful and insightful discussions.

A.R. acknowledges funding from the Swiss National Science Foundation (Postdoc.Mobility grant `CATCH' P500PT225461) and from the European Research Council (Consolidator grant ‘Cocoquest’ 101043705). P.A. acknowledges fundings from the Austrian Science Fund (FWF) projects I-6004 and ESP2889224. P.L.-B. acknowledges funding from the Polish National Science Centre through project Sonata 2023/51/D/ST2/02309.  M.P.-L.
acknowledges support from the Grant ATR2024-154621
funded by the Spanish MICIU/AEI/10.13039/501100011033. M.A. and P.J.C. acknowledge funding from the Advanced Research and Invention Agency’s
(ARIA) Scaling Compute programme.

\bibliography{refs}
\appendix

\widetext
\section{Quadratic Systems}
\label{app:physics}
\subsection{Dynamics}
We consider a system of $N$ Brownian particles described by the position vector $x$ and interacting using an arbitrary quadratic time-dependent potential $V(x, t)$, that is
\begin{align}
    \label{eq:quadratic_potential}
    V(x, t) = \frac{1}{2} x^T K(t) x ~.
\end{align}
The symmetric and positive semi-definite matrix $K$ describes the interactions between the particles. The systems' dynamics is governed by the potential $V(x, t)$, as well as a random stochastic force $\eta$ modeled as Gaussian white noise, i.e. $\langle \eta \rangle = 0$ with $\langle \eta_i(t) \eta_j(t')\rangle = \delta_{ij} \delta(t-t')$. As a consequence, we describe the system evolution using a set of overdamped Langevin equations of the form
\begin{align}
    \label{eq:langevin}
    \dot{x} = -\frac{1}{\gamma} \nabla_{x} V(x, t) + \sqrt{\frac{2}{\gamma \beta}} \eta = -\frac{1}{\gamma} K(t) x + \sqrt{\frac{2}{\gamma \beta}}\eta. 
\end{align}
We will assume that the system initially starts in a Gaussian state at time $t=0$, hence it remains Gaussian at all times $t > 0$. Consequently, its state can and can be described using a probability density function:
\begin{align}
    \rho(x, t) = \frac{1}{\sqrt{(2\pi)^N \det \Sigma(t)}} \exp(-\frac{1}{2} x^{T} \Sigma(t)^{-1}x)~,
\end{align}
where the covariance matrix $\Sigma(t) = \langle x x^{T} \rangle(t)$, which provides a full description of the state at time $t$. It is worth noting that this PDF has zero mean $\langle x\rangle = 0$, however it can be easily generalized to a case where the mean of the distribution is non-trivial. We can obtain a differential equation for the covariance matrix by observing that $\dot{\Sigma} = \langle x\dot{x}^{T}\rangle + \langle \dot{x}x^{T}\rangle$ and using \eref{eq:langevin}
\begin{align}
\label{eqapp:langevin_sigma}
    \dot{\Sigma}(t) = - K(t) \Sigma(t) - \Sigma(t) K(t) + \frac{2}{\beta} \mathbb{1}~, 
\end{align}
where we set $\gamma=1$.
For a given control trajectory $K(t)$ this equation is solved by
\begin{align}\label{eq:evolution}
    \Sigma(t) = e^{-\int_0^t\!dsK(s)}\Sigma(0)e^{-\int_0^t\!dsK(s)} + \frac{2}{\beta}\int_0^t\!ds~e^{-2\int_s^t\!drK(r)}~.
\end{align}
If instead we consider a fixed state trajectory $\Sigma(t)$, we can solve for the control $K(t)$ that realizes that trajectory
\begin{align}\label{eq:control}
    K(t) = \beta^{-1}\Sigma(t)^{-1} - \int_{0}^{\infty}\! d\nu~ e^{-\nu \Sigma(t)} \dot{\Sigma}(t) e^{-\nu \Sigma(t)}.
\end{align}
The expressions we obtained here will be used later when we will look for optimal protocols. Moreover, we will refer to $K(t)$ as a finite-time protocol that is realized in some time $\tau > 0$. Given that for $t>\tau$ the control remains constant, we can see from \eref{eq:evolution} that the state converges to its thermal state, with covariance matrix $\Sigma_{th} = \beta^{-1}K(\tau)^{-1}$.

\subsection{Thermodynamics}
The average internal energy of a system in the potential from \eref{eq:quadratic_potential} is given by
\begin{align}
    E(t) = \int\!dx~\rho(x, t) V(x, t)  = \frac{1}{2} \sum_{i,j} K_{ij}(t) \langle x_i x_j \rangle (t) = \frac{1}{2} \Tr[K(t) \Sigma(t)]~.
\end{align}
In stochastic thermodynamics the work $\dot{W}$ and heat $\dot{Q}$ \emph{fluxes} of the system are traditionally defined as~\cite{seifert2012stochastic}:
\begin{align}
    \dot{W}(t) := \frac{1}{2} \Tr[\dot{K}(t) \Sigma(t)], \qquad \dot{Q}(t) := \frac{1}{2} \Tr[K(t) \dot{\Sigma}(t)].
\end{align}
Integrating the work flux over time and using \eref{eq:control} we find
\begin{align}
    W &= \frac{1}{2} \int_{0}^{\tau}\!dt~  \Tr[\dot{K}(t) \Sigma(t)]~, \\
    &= \frac{1}{2} \Tr[K(t) \Sigma(t)]\Big|_{0}^{\tau} - \frac{1}{2 \beta} \ln \det \Sigma(t) \Big|_{0}^{\tau} + \frac{1}{2} \int_{0}^{\tau}\!dt\Tr[\int_{0}^{\infty}\!d\nu~ e^{-\nu \Sigma(t)} \dot{\Sigma}(t) e^{-\nu \Sigma(t)} \dot{\Sigma}(t) ]~.
\end{align}
Now let us make two observation. First, note that that the quantity 
\begin{align}
\frac{1}{2} \ln \det \Sigma(t) \Big|_{0}^{\tau} = - \left[\int\!dx~ \rho(x, t) \ln \rho(x, t) \right]_0^{\tau} = \Delta S    
\end{align}
is the change of the von Neumann entropy of the system. Second, observe that the term 
\begin{align}
    \frac{1}{2} \Tr[K(t) \Sigma(t)]|_{0}^{\tau} = E(\tau) - E(0) = \Delta E,
\end{align}
is the energy change of the system during the protocol. Therefore we can identify the first to terms as the difference in non-equilibrium free energy ($F = E-\beta^{-1}S$). This leads to the last term corresponding to the dissipated work $W_{\rm diss} := W - \Delta F$
\begin{align}
    \label{eq:w_irr}
    W_{\rm diss} =  \frac{1}{2} \int_{0}^{\tau}\!dt \Tr[\int_{0}^{\infty} e^{-\nu \Sigma(t)} \dot{\Sigma}(t) e^{-\nu \Sigma(t)} \dot{\Sigma}(t) \,\,\text{d} \nu ].
\end{align}
The above integral is precisely the integral over time of a Bures-Wasserstein (BW) quadratic form. More specifically, by defining $\B\W(\Sigma, \Sigma + d \Sigma)^2 := g_{\Sigma}(d \Sigma, d \Sigma)$ with
\begin{align}
    g_{\Sigma}(A, B) = \frac{1}{2} \int_{0}^{\infty} \Tr[e^{-\nu \Sigma} A e^{- \nu \Sigma} B] \, d \nu,
\end{align}  
one can express the entropy production from \eref{eq:w_irr} as
\begin{align}
\label{eqapp:Wdiss_metric}
    W_{\rm diss} = \int_{0}^{\tau} g_{\Sigma} (\dot{\Sigma}, \dot{\Sigma}) \, d t.
\end{align}
Consider now a finite-time protocol with endpoints $\Sigma_0 = \Sigma(0)$ and $\Sigma_1 = \Sigma(\tau)$. For such process we can obtain an achievable lower bound on $W_{\rm diss}$ by computing the geodesic with respect to the Bures-Wasserstein quadratic form. In the current context this is given by
\begin{align}
    W_{\text{diss}} \geq \frac{\B\W(\Sigma_0, \Sigma_1)^2}{\tau},
\end{align}
where $\B\W$ is the Bures-Wasserstein distance~\cite{bengtsson2017geometry} defined as
\begin{align}\label{eq:distance}
    \B\W(\Sigma_0, \Sigma_1)^2 := \Tr[\Sigma_0] + \Tr[\Sigma_1] - 2\Tr[(\Sigma_0^{1/2} \Sigma_1 \Sigma_0^{1/2})^{1/2}].
\end{align}
For our use-case, it is worth noting that we can freely choose $\Sigma_0 \propto\mathbb{1}$, and therefore it is useful to consider \eref{eq:distance} when $[\Sigma_0,\Sigma_1] = 0$. In particular, if we denote by $\sigma^{(j)}_i$ the eigenvalues of $\Sigma_i$ we find that the distance splits as a sum over the eigenvalues of the matrices
\begin{equation}\label{eq:separation_work}
    \B\W(\Sigma_0,\Sigma_1)^2 = \sum_\alpha \left(\sqrt{\sigma_{0,\alpha}} - \sqrt{\sigma_{1,\alpha}}\right)^2~.
\end{equation}

\section{Symmetries in Driving Protocols}
\label{app:symmetries}
The design of EDP‐optimal driving protocols can be simplified by using two complementary symmetries of the entropy production $W_{\text{diss}}$, namely {temperature rescaling} which increases linearly the entropy production, and {time-strength} rescaling which leaves the EDP invariant.

\smallskip 
\noindent \textbf{Temperature rescaling.}  
From the evolution equation for the covariance matrix \(\Sigma(t)\) (\eref{eq:evolution}) one sees that \(\Sigma(t)\propto\beta^{-1}\) (since the initial state is a thermal distribution).  Therefore, the transformation $\beta^{-1}\longrightarrow\lambda\beta^{-1}$ leads to
\begin{align}
\Sigma(t)\;\longrightarrow\;\lambda\Sigma(t), \qquad
\dot\Sigma(t)\;\longrightarrow\;\lambda\dot\Sigma(t),
\end{align}
which, upon substitution into the definition of irreversible work (\eref{eq:w_diss_quad}), yields
\begin{equation}
W_{\text{diss}}
\;\longrightarrow\;
\frac{\lambda^2}{2}\!\!
\int_{0}^{\tau}\!dt\Tr[\int_{0}^{\infty}\!d\nu~
e^{-\nu\,\lambda\,\Sigma(t)}\,\dot\Sigma(t)\,
e^{-\nu\,\lambda\,\Sigma(t)}\,\dot\Sigma(t)]
\;=\;\lambda\,W_{\text{diss}}~,
\end{equation}
where we used a change in variable on $\nu$.
Thus, \(W_{\rm diss}\) is \emph{linear} in the temperature \(\beta^{-1}\).

\smallskip 
\noindent \textbf{Time–strength rescaling.} 
Consider a finite‐time protocol \(K(t)\) over \(t\in[0,\tau]\).  The transformation
\begin{equation} \label{eq:lie_action}
K(t)\;\longrightarrow\;K'(t)=\lambda K(\lambda t)~,\qquad
t\in\bigl[0,\tau/\lambda\bigr]~,
\end{equation}
simultaneously amplifies the control strength by \(\lambda\) and compresses its duration to \(\tau/\lambda\).  One then finds that the new covariance $\Sigma'(t)$ evolves as
\begin{align}
\Sigma'(t)
&=e^{-\int_{0}^{t}\!\text{d}s\,\lambda K(\lambda s)}\,\lambda^{-1}\Sigma(0)\,e^{-\int_{0}^{t}\!\text{d}s\,\lambda K(\lambda s)}
+\frac{2}{\beta}\!\int_{0}^{t}\!\text{d}s\,e^{-2\int_{s}^{t}\!\text{d}r\,\lambda K(\lambda r)} = \lambda^{-1}\Sigma(\lambda t)~.
\end{align}
This leads to the following rescaling 
\begin{align}
    \Sigma(t) \;\longrightarrow\; \lambda^{-1}\,\Sigma\bigl(\lambda t\bigr)~,
\qquad 
\dot\Sigma(t) \;\longrightarrow\; \dot\Sigma\bigl(\lambda t\bigr)~.
\end{align}
Substituting this into \eref{eq:w_diss_quad} shows
\begin{equation}
W_{\text{diss}}
\;\longrightarrow\;
\frac{1}{2}\!\!\int_{0}^{\tau/\lambda}\!dt\Tr[\int_{0}^{\infty}\!d\nu~
e^{-\nu\,\lambda^{-1}\Sigma(\lambda t)}\,\dot\Sigma(\lambda t)\,
e^{-\nu\,\lambda^{-1}\Sigma(\lambda t)}\,\dot\Sigma(\lambda t)]
= W_{\text{diss}}~,
\end{equation}
so \(W_{\text{diss}}\) is invariant under the action from \eref{eq:lie_action}. This symmetry implies that amplifying the protocol’s strength by $\lambda$ while compressing its duration to $\tau/\lambda$ yields an equivalent thermodynamic outcome. Formally, the transformations form the Lie group $(\mathbb{R}^+, \cdot)$, acting on the space of protocols $\mathcal{F}$. The orbit of a protocol $K(t)$, given by $\{ \lambda K(\lambda t) \mid \lambda \in \mathbb{R}^+ \}$, defines an equivalence class with constant EDP. This scaling symmetry reduces the optimization problem by allowing us to search for optimal protocols within the quotient space $\mathcal{F}/\mathbb{R}^+$.

\smallskip 
\noindent \textbf{Implications for optimization.}  The time-strength symmetry \(K(t)\to\lambda K(\lambda t)\) partitions the space of protocols \(\mathcal{F}\) into orbits (equivalence classes) of constant irreversible work.  Hence one can restrict the search for EDP‐optimal protocols to the quotient space \(\mathcal{F}/\mathbb{R}^+\).  Finally, both of the discussed symmetries  for entropy production also hold for the Bures-Wasserstein distance, as it is realized by the geodesic protocol.

\section{Further Details of Protocols}
\label{app:protocols}

\subsection{Quenches and Bang-Bang Protocols}
In a quench protocol (see Ref.~\cite{aifer2024_TLA}) the system potential is driven instantaneously from $V_0(x)$ to $V_1(x)$, meaning that formally $\tau_D = 0$. The process can therefore be written as
\begin{align}
    (\Sigma_0, K_0) \xrightarrow[(a)]{\text{quench}} (\Sigma_0, K_1) \xrightarrow[(b)]{\text{thermalization}} (\Sigma_1, K_1).
\end{align}
In sub-process $(a)$ we have:
\begin{align}
    W_{a} = \frac{1}{2} \tr[(K_1 - K_0)\Sigma_0], \qquad \Delta E_{a} = \frac{1}{2} \Tr[(K_1 - K_0) \Sigma_0], \qquad \Delta S_{a} = 0. 
\end{align}
In sub-process $(b)$ we have:
\begin{align}
    W_{b} = 0, \qquad \Delta E_b = \frac{1}{2} \Tr[K_1(\Sigma_1 - \Sigma_0)], \qquad \Delta S_b = \frac{1}{2}(\ln \det \Sigma_1 - \ln \det \Sigma_2).
\end{align}
Therefore the entropy production during the whole protocol composed of steps $(a)$ and $(b)$ is given by
\begin{align}
    W_{\text{diss}}^Q = \sum_{x \in \{a, b\}} W_x - (\Delta E_x - T \Delta S_x) = \frac{1}{2} \Tr [K_1(\Sigma_0 -\Sigma_1)] + \frac{1}{2}T \ln \left(\frac{\det \Sigma_1}{\det \Sigma_0}\right),
\end{align}
then we simply take $\Sigma_0$ and $\Sigma_1$ to be their respective thermal states.

Using this same approach we can consider Bang-Bang protocols, where a quench is performed to an intermediary control $K_*$ at $t=0$ and a second quench to $K_1$ is performed at $t=\tau_D/2$
\begin{equation}
    W_{\text{diss}}^{\text{Bang-Bang}} = W_{\text{diss}}^{Q} - \frac{1}{2}\Tr [(K_1-K_*)(TK_0^{-1} - \Sigma_*)]~,
\end{equation}
where we can compute $\Sigma_*$ with \eref{eq:evolution} 
\begin{equation}
    \Sigma_* = e^{-K_*\tau_D/2}\Sigma_0 e^{-K_*\tau_D/2} + (\mathbb{1}-e^{-K_*\tau_D})TK_*^{-1}~.
\end{equation}

\subsection{Geodesics}

Generally, all protocols will dissipate some energy into the environment. It can be shown that for the system at hand, we can describe it as a property of the length of a trajectory in the state space equipped with the Bures-Wasserstein metric~\cite{abiuso2022thermodynamics}. This implies that we can minimize it with a protocol that matches geodesics in this space
\begin{equation}    
\Sigma_{\text{geo}}(t) = (1-s)^2\Sigma_0 + s^2\Sigma_1 + s(1-s)(\sqrt{\Sigma_0 \Sigma_1} + \sqrt{\Sigma_1 \Sigma_0})
\end{equation}
where $s=t/\tau_D$.
The state, the corresponding control that realizes it is obtained by inserting the geodesic state trajectory into \eref{eq:control}
\begin{equation}\label{eq:opt_path}
    K(t) = T\Sigma_{\text{geo}}(t)^{-1} - \frac{1}{\tau_D}\left(s\Sigma_1 + (1-2s)\sqrt{T\Sigma_1}-(1-s)T\mathbb{1}\right)\Sigma_{\text{geo}}(t)^{-1}~,
\end{equation}
for $0<t<\tau_D$. It is worth noting that since at the extremities of the protocol we have $K(0) = T\Sigma_0^{-1}$ and $K(\tau) = T\Sigma_1^{-1}$ \eref{eq:opt_path} implies that there will be jumps at the beginning and end of the protocol:
\begin{align}
    K(0^+)-K(0) &= \frac{1}{\tau}\left(\mathbb{1}- \sqrt{\Sigma_1/\Sigma_0}\right)~,\\
    K(\tau)-K(\tau^-) &= \frac{1}{\tau}\left(\mathbb{1}- \sqrt{\Sigma_0/\Sigma_1}\right)~.
\end{align}
This protocol will dissipate 
\begin{equation}
    W_{\rm diss}^{(\text{opt})} = \frac{\B\W(\Sigma_0, \Sigma_1)^2}{\tau_D}~.
\end{equation}

\subsection{Slow driving protocols}
The metric featured in \eref{eqapp:Wdiss_metric} is dependent on the state of the system (encoded in $\Sigma(t)$) instead of the externally-controlled $K(t)$. This implies that for a given protocol on the system, the dissipated work depends on the exact solution of the dynamics~\eq{eqapp:langevin_sigma}. However, in the slow driving limit~\cite{abiuso2022thermodynamics} (at first order) this dependence greatly simplifies. In this limit we only consider the control over the trajectory. We can simplify the analysis of comparing different protocols by restricting ourselves to the slow-driving regime and expanding \eref{eqapp:Wdiss_metric} via $\Sigma(t)=K^{-1}(t)+\mathcal{O}(\tau^{-1})$, as
\begin{align}
    W_{\rm diss} = \int_{0}^{\tau} g_{K^{-1}} (K^{-1}\dot{K}K^{-1}, K^{-1}\dot{K}K^{-1}) \, d t\; +\mathcal{O}(\tau^{-2})
\end{align}
This expression further simplifies in the commuting case
$[K_1,K_0]=0$,
\begin{equation}
    W_{\rm diss} = \frac{1}{4}\int_0^\tau \Tr\!\left[\frac{\dot K(t)^2}{K(t)^3}\right] + \mathcal{O}(\tau^{-2})~,
\end{equation}
or equivalently with the eigenvalues of $K$
\begin{align}
    W_{\rm diss} =  \frac{1}{4} \int_{0}^{\tau}\!dt\sum_i \frac{\dot{\kappa}_i^2}{\kappa_i^3}\;.
\end{align}
If we decompose the boundary conditions in terms of their eigenvalue $k^{(i)}$ we obtain
\begin{equation}
\label{eq:Wirr_fk}
    W_{\rm diss} = \sum_{i} f(k^{(0)}_{i},k^{(1)}_i)~,
\end{equation}
with $f(x,y) = \frac{T(x-y)^2}{4}\int_0^\tau\!dt \frac{\dot s(t)^2}{((1-s(t))x + s(t)y)^3}$. 
If the matrices $K_0$ and $K_1$ are sampled from some distribution, the average $\langle W_{\rm diss}\rangle$, as well as its fluctuations, will depend on it. Here let us make us two observations. Assume for simplicity that $K_0\propto \mathbb{1} $, therefore $k_i^{(0)}\equiv \bar{\kappa}$. We have
\begin{align}
    \langle  W_{\rm diss}\rangle=\sum_{i} \langle f(\bar{\kappa},k^{(1)}_i)\rangle
\end{align}
When the sample distribution of $K_1$ is isotropic, then we expect the eigenvalues to be identically distributed. Which implies that the expectation value of the dissipation is given by 
\begin{align}
    N\langle f(\bar{\kappa},k_1)\rangle\;.
\end{align}
In general we do not have explicit versions of $f(x,y)$, however for some cases it is simple enough that we can use the expression to compute expectation values. For example, in the case of a linear ramp we have $f(x,y) = \frac{T(x-y)^2(x+y))}{8x^2y^2}$. 
Secondly, if the eigenvalues $k_i$ are also uncorrelated, we will have
\begin{align}
    {\rm Var}(W_{\rm diss})=N{\rm Var}(f(\bar{\kappa},k_1))
\end{align}
which explains the concentration of  $W_{\rm diss}$ for large $N$. Notice that the same happens whenever the eigenvalues $k_i$ are not "fully correlated", i.e. whenever ${\rm Var}[\sum_i f(\bar{\kappa},k_i)]=\mathcal{O}(N^{\alpha})$ with $\alpha< 2$.

\section{Natural squeezing and partial optimization of \texorpdfstring{$W^{\rm(opt)}_{\rm diss}$}{dissipated work}}
\label{app:scale_optimization}
Consider the standard case in which either the initial or the final covariance matrix (and associated quadratic potential) is isotropic, e.g. boundary conditions of the form 
\begin{align}
    \Sigma_0=\alpha\mathbb{1}\;,\quad \Sigma_1=K^{-1}\;,
\end{align}
to which is associated a minimum dissipation given by the $\B\W$ distance~\eq{eq:distance}-\eq{eq:separation_work} 
\begin{align}
    \label{eqapp:alpha_start}\B\W^2=\Tr[(\sqrt{\alpha}\mathbb{1}-K^{-1/2})^2]\;.
\end{align}
Assuming, e.g.,  that the initial (idle-mode) squeezing $\alpha^{-1}$ is under experimental control, we will now provide useful bounds to the minimization of $\B\W$. Similar considerations follow when the squeezing of $K$ can be tuned.

\subsection{Tuning \texorpdfstring{$\alpha$}{alpha}}
When being able to tune $\alpha$, the minimum of the quadratic expression~\eq{eqapp:alpha_start} is obtained for $\alpha^*$ satisfying
\begin{align}
    N\sqrt{\alpha^*}=\Tr[K^{-1/2}]\;,
\end{align}
$N$ being the dimension of the matrices we are considering.
Clearly we cannot, in general, assume to know $\Tr[K^{-1/2}]$, as this would need diagonalizing/inverting the matrix $K$. We can however provide lower bounds which then serve as a rule of thumb for choosing $\alpha$.

For example from the mean-inequalities we know that for positive eigenvalues $\lambda_i$ it holds
\begin{align}
    \left(\frac{1}{N}\sum_i \lambda_i^{-1/2}\right)^{-2}\leq \left(\prod_i \lambda_i\right)^{\frac{1}{N}} \leq \frac{1}{N}\sum\lambda_i\;.
\end{align}
The same inequality can thus be re-expressed as
\begin{align}
    \alpha^*\geq (\det K)^{-\frac{1}{N}}\geq \frac{N}{\Tr[K]}\;.
\end{align}
Notice that it is only possible to \emph{lower} bound $\alpha^*$ without using "expensive" calculations, as there could always by eigenvalues of $K$ small enough to make the optimal $\alpha^*$ increase.
However, if we include semi-definite-programming (SDPs) in the set of functions that are "not expensive" to compute, we can access to $\lambda_{\rm min}$ of $K$ and bound
\begin{align}
    \alpha^*\leq \frac{1}{\lambda_{\rm min}}\;.
\end{align}
\subsection{Fixed \texorpdfstring{$\alpha$}{alpha}, rescaling \texorpdfstring{$K$}{K}}
Suppose now that $\alpha$ is fixed, but we have the freedom to rescale $K$ (this corresponds to a case in which one does not have experimental control over the squeezing of the idle-mode of the thermodynamics machine, but is free to remap the problem of interest by rescaling the coordinates of $\rho(x)$). We thus consider
\begin{align}
    \B\W^2=\Tr[(\sqrt{\alpha}\mathbb{1}-\eta^{-1/2}K^{-1/2})^2]
\end{align}
and vary $\eta$. One can easily see that the minimum is obtained for 
\begin{align}
    \sqrt{\alpha}\Tr[K^{-1/2}]={\eta^*}^{-1/2}\Tr[K^{-1}]\;.
\end{align}
Once again, one can use inequalities to give bounds to $\eta^*$.
For example noticing
\begin{align}
    \frac{\Tr[K^{-1}]}{\Tr[K^{-1/2}]}\leq \lambda_{\rm min}^{-1/2}\;,
\end{align}
one gets
\begin{align}
    \eta^* \leq \frac{\lambda_{\rm min}^{-1}}{\alpha}\;.
\end{align}
Similarly, via a Cauchy-Schwarz inequality we have
\begin{align}
    \frac{\Tr[K^{-1}]}{\Tr[K^{-1/2}]}\geq \frac{{\Tr[K^{-1/2}]}}{N}\geq \frac{\sqrt{N}}{\sqrt{\Tr[K]}}\;,
\end{align}
and therefore
\begin{align}
    \eta^*\geq \frac{N}{\alpha\Tr[K]}\;.
\end{align}
Notice that these inequalities obtained for $\eta^*$ are dual to those for $\alpha^*$ above.

\end{document}